\RequirePackage[compress]{cite} 
\documentclass[final,3p,linenumbers,compress]{elsarticle}

\makeatletter
\newcommand*{\rom}[1]{\expandafter\@slowromancap\romannumeral #1@}
\makeatother

\usepackage{lineno,hyperref}
\usepackage{multirow}

\usepackage{amssymb}
\usepackage{amsmath}
\usepackage{graphicx,color}

\usepackage{cite}

\usepackage{multicol}

\usepackage{float}

\usepackage{xspace}	

\newcommand{\pt}           {\ensuremath{p_{\rm T}}\xspace}

\newcommand{\roots} {\mbox{$\sqrt{\textit{s}_{NN}}$}\xspace}
\newcommand{\GeVc} {\mbox{GeV/$\textit{c}$}\xspace}

\def  \vn          {\mbox{$\textit{v}_{n}$}\xspace} 
\def  \first       {\mbox{$\textit{v}_{1}$}\xspace}
\def  \second      {\mbox{$\textit{v}_{2}$}\xspace}
\def  \third       {\mbox{$\textit{v}_{3}$}\xspace}
\def  \fourth      {\mbox{$\textit{v}_{4}$}\xspace}
\def  \fifth       {\mbox{$\textit{v}_{5}$}\xspace}
\def  \higher      {\mbox{$\textit{v}_{n}$}\xspace ($n > 3$)}
\def  \etas        {\mbox{$\eta / \textit{s}$  }\xspace}

\def \sc23  {\mbox{$\mathrm{SC}(2,3)$   }\xspace}
\def \sc24  {\mbox{$\mathrm{SC}(2,4)$   }\xspace}

\def \nsc23 {\mbox{$\mathrm{NSC}(2,3)$}\xspace}
\def \nsc24 {\mbox{$\mathrm{NSC}(2,4)$}\xspace}

\journal{Phys. Lett. B}

\DeclareUnicodeCharacter{2212}{-}

\begin{document}


\begin{frontmatter}

\title{Beam energy dependence of the linear and mode-coupled flow harmonics in Au+Au collisions}

\author{
B.~E.~Aboona$^{52}$,
J.~Adam$^{15}$,
J.~R.~Adams$^{38}$,
G.~Agakishiev$^{28}$,
I.~Aggarwal$^{39}$,
M.~M.~Aggarwal$^{39}$,
Z.~Ahammed$^{57}$,
A.~Aitbaev$^{28}$,
I.~Alekseev$^{2,35}$,
D.~M.~Anderson$^{52}$
% ...
}

\address{\rm{(STAR Collaboration)}}

% ...
\address{$^{2}$Alikhanov Institute for Theoretical and Experimental Physics NRC "Kurchatov Institute", Moscow 117218}
% ...
\address{$^{15}$Czech Technical University in Prague, FNSPE, Prague 115 19, Czech Republic}
% ...
\address{$^{28}$Joint Institute for Nuclear Research, Dubna 141 980}
% ...
\address{$^{35}$National Research Nuclear University MEPhI, Moscow 115409}
% ...
\address{$^{38}$Ohio State University, Columbus, Ohio 43210}
\address{$^{39}$Panjab University, Chandigarh 160014, India}
% ...
\address{$^{52}$Texas A\&M University, College Station, Texas 77843}
% ...
\address{$^{57}$Variable Energy Cyclotron Centre, Kolkata 700064, India}
% ...



\begin{abstract}
The linear and mode-coupled contributions to higher-order anisotropic flow are presented for Au+Au collisions at \roots = 27, 39, 54.4, and 200~GeV and compared to similar measurements for Pb+Pb collisions at the Large Hadron Collider (LHC). The coefficients and the flow harmonics' correlations, which characterize the linear and mode-coupled response to the lower-order anisotropies, indicate a beam energy dependence consistent with an influence from the specific shear viscosity ($\eta/s$). In contrast, the dimensionless coefficients, mode-coupled response coefficients, and normalized symmetric cumulants are approximately beam-energy independent, consistent with a significant role from initial-state effects. These measurements could provide unique supplemental constraints to (i) distinguish between different initial-state models and (ii) delineate the temperature ($T$) and baryon chemical potential ($\mu_{B}$) dependence of the specific shear viscosity $\frac{\eta}{s} (T, \mu_B)$. 
\end{abstract}



\end{frontmatter}

\begin{multicols}{2}

Experimental studies of heavy-ion collisions at the LHC and the Relativistic Heavy Ion Collider (RHIC) indicate the creation of the Quark-Gluon Plasma (QGP)~\cite{Shuryak:1978ij,Shuryak:1980tp,Muller:2012zq,STAR:2000ekf}, a state of matter predicted by Quantum Chromodynamics (QCD). A central aim of prior and current experimental investigations of this plasma is to understand its transport properties such as its specific viscosity or ratio of shear viscosity to entropy density (\etas)~\cite{Shuryak:2003xe,Romatschke:2007mq,Luzum:2008cw,Bozek:2009dw,Acharya:2019vdf,Acharya:2020taj,Adam:2020ymj}. Anisotropic flow measurements continue to be a valuable route to \etas estimation because they reflect the viscous hydrodynamic response to the anisotropy of the initial-state energy density~\cite{Heinz:2001xi,Hirano:2005xf,Huovinen:2001cy,Hirano:2002ds,Romatschke:2007mq,Luzum:2011mm,Song:2010mg,Qian:2016fpi,Magdy:2017ohf,Magdy:2017kji,Schenke:2011tv,Teaney:2012ke,Gardim:2012yp,Lacey:2013eia}
which is characterized by the complex eccentricity vectors $\mathcal{E}_{n}$~\cite{Alver:2010dn,Petersen:2010cw,Lacey:2010hw,Teaney:2010vd,Qiu:2011iv}:
\begin{eqnarray}
\mathcal{E}_{n} & \equiv &  \varepsilon_{n} e^{i {\textit{n}} \Phi_{n} }  \\  \nonumber
&\equiv  &
  - \frac{\int dx\,dy\,\textit{r}^{n}\,e^{i {\textit{n}} \phi}\, \rho_e(r,\phi)}
           {\int dx\,dy\,\textit{r}^{n}\,\rho_e(r,\phi)}, ~(\textit{n} ~>~ 1),
\label{epsdef1}
\end{eqnarray}
where  $\varepsilon_{n}$ and $\mathrm{\Phi_{n}}$ are the magnitude and azimuthal direction of the n$^{\rm th}$ eccentricity vector, $x=r~\cos\phi$, $y=r~\sin\phi$, $r$ is the radial coordinate, 
 $\phi$ is the spatial azimuthal angle, and $\rho_e(r,\phi)$ is the initial energy density profile~\cite{Teaney:2010vd,Bhalerao:2014xra,Yan:2015jma}. 

{\color{black}The azimuthal anisotropy of particles produced can be expressed as~\cite{Poskanzer:1998yz}:}
 {\small
 \begin{equation}
\label{eq:1-1}
E_{p} \frac{d^{3}N}{d^3p} =  \dfrac{1}{2\pi} \dfrac{d^2N}{p_{T} dp_{T} dy} 
                                             \left( 1 +  \sum^{\infty}_{n=1} 2 \textit{v}_{n}  \cos\left(    n \left( \varphi - \psi_{n} \right)    \right)        \right),
\end{equation} }
where $N$ is the number of the particles produced, $E_{p}$ is the energy of the particle, $p_{T}$ is transverse momentum, $y$ is the rapidity,  and $\varphi$ is the azimuthal angle of the particle's momentum; {\color{black} \vn and $\psi_{n}$ represent the magnitude and the direction of the vector $V_{n} = v_{n} e^{i n \psi_{n}}$.}
%
The coefficients  \first, \second, and \third are commonly termed directed, elliptic and triangular flow, respectively.

Prior investigations of $v_2$ and $v_3$ and their fluctuations~\cite{STAR:2018fpo,ALICE:2016kpq,Adamczyk:2017hdl,Qiu:2011iv, Adare:2011tg, Aad:2014fla, Aad:2015lwa,Magdy:2018itt,Alver:2008zza,Alver:2010rt, Ollitrault:2009ie,Magdy:2022jai,STAR:2022gki} as well as higher-order flow harmonics \higher~\cite{Magdy:2019ojv,Adam:2019woz,Magdy:2018itt,Adamczyk:2017ird,Magdy:2017kji,Adamczyk:2017hdl,Alver:2010gr, Chatrchyan:2013kba,Magdy:2022cvt} have provided invaluable initial insights into the properties of the QGP. Notably, the extensively studied \second~\cite{Adamczyk:2016gfs,Adamczyk:2015fum,Magdy:2018itt,Adamczyk:2015obl} and \third flow coefficients~\cite{Adamczyk:2016exq,Adam:2019woz} are linearly related to $\varepsilon_{{{2}}}$ and $\varepsilon_{{{3}}}$~\cite{Song:2010mg, Niemi:2012aj,Gardim:2014tya,Fu:2015wba,Holopainen:2010gz,Qin:2010pf,Qiu:2011iv,Gale:2012rq,Liu:2018hjh,Magdy:2022ize}:
\begin{eqnarray}
\label{eq:1-2}
v_{n} &=& \kappa_{n} \varepsilon_{n},
\end{eqnarray}
where the parameter $\kappa_{n}$ encodes the effects of viscous attenuation~\cite{Liu:2018hjh,Adam:2019woz,Heinz:2013th} which depend on the particle $p_T$, charged particle multiplicity and \etas.
The higher-order flow harmonics show a linear response to the same-order eccentricity but also include a mode-coupled response to the lower-order eccentricities $\varepsilon_{{{2}}}$ and $\varepsilon_{{{3}}}$~\cite{Teaney:2012ke,Bhalerao:2014xra,Yan:2015jma,Gardim:2011xv}:
\begin{eqnarray}
V_{4}  &=&  v_{4} e^{i 4 \psi_{4}} = \kappa_{4} \varepsilon_{4} e^{4i\Phi_{4}} + \kappa_{4}^{'} \varepsilon^{2}_{2} e^{4i\Phi_{2}}
\nonumber \\
           &=&  V_{4}^{\rm Linear} +   \chi_{4,22} V_{4}^{\rm MC}, \label{eq:1-3a} \\
V_{5}  &=&  v_{5} e^{i 5 \psi_{5}} = \kappa_{5} \varepsilon_{5} e^{5i\Phi_{5}} + \kappa_{5}^{'} \varepsilon_{2} e^{2i\Phi_{2}} \varepsilon_{3} e^{3i\Phi_{3}} \nonumber \\
       &=& V_{5}^{\rm Linear} +   \chi_{5,23} V_{5}^{\rm  MC}, \label{eq:1-3b}
\end{eqnarray}
where $\kappa_{k}^{'}$ ($k=4,5$) reflects the combined influence of the medium properties and the coupling between the lower- and higher-order eccentricity harmonics. In Eqs.~(\ref{eq:1-3a}) and (\ref{eq:1-3b}) the  terms $\textit{V}^{\rm Linear}_{k}$ and $\textit{V}^{\rm MC}_{k}$ are the linear and the mode-coupled contributions and $\chi_{k,nm}$ represents the mode-coupled response coefficients. 

The mode-coupled contributions to $\textit{V}_{k}$ and the normalized symmetric cumulants $\mathrm{NSC (n,m)}$ can provide further constraints for $\eta/s$ and the initial-stage dynamics~\cite{Bilandzic:2013kga, Bhalerao:2014xra, Aad:2015lwa, ALICE:2016kpq, STAR:2018fpo,Zhou:2016eiz, Qiu:2012uy,Teaney:2013dta, Niemi:2015qia, Zhou:2015eya}. Consequently, ongoing efforts seek to leverage extensive measurements of the linear and mode-coupled contributions to $\textit{V}_{k}$ and $\mathrm{NSC (n,m)}$ to develop unique supplemental constraints that can (i) distinguish between different initial-state models and (ii) pin down the temperature ($T$) and baryon chemical potential ($\mu_{B}$) dependence of the specific shear viscosity $\frac{\eta}{s} (T, \mu_B)$; note that $T$ and $\mu_B$ vary with beam energy. Prior measurements have been reported for charged hadrons in Pb+Pb collisions at \roots = 2.76 and 5.02~TeV \cite{Acharya:2017zfg,ALICE:2020sup,CMS:2019nct} and Au+Au collisions at \roots = 200~ GeV \cite{Adam:2020ymj,STAR:2018fpo}, and for identified particle species in Pb+Pb collisions at \roots = 2.76~GeV \cite{Acharya:2017zfg}. Here, we report the $\textit{V}^{\rm Linear}_{n}$, $\textit{V}^{\rm MC}_{n}$, $\chi_{k,nm}$ and $\mathrm{NSC (n,m)}$ measurements for  Au+Au collisions at \roots = 27, 39, 54.4, and 200~GeV to extend the data set that can provide simultaneous constraints for $\frac{\eta}{s} (T, \mu_B)$ and the initial-state. The initial-state effects which influence the dimensionless mode-coupled coefficients and the normalized symmetric cumulants could be insensitive to the beam energy, while $\frac{\eta}{s} (T, \mu_B)$ is not~\cite{Lacey:2013qua,Karpenko:2015xea,Magdy:2021sba}.


The data for the present analysis were collected with the STAR detector at RHIC using a minimum-bias trigger~\cite{Judd:2018zbg} in 2017, 2010 and 2018 at \roots = 54.4, 39 and 27~GeV respectively. Charged particle tracks with full azimuthal angle and pseudorapidity $|\eta|<1.0$ coverage were used to reconstruct the collision vertices of tracks measured in the Time Projection Chamber (TPC)~\cite{Anderson:2003ur}. A Monte Carlo Glauber simulation has been used to determine the collision centrality from the measured event-by-event charged particle multiplicity in $|\eta|<0.5$ with at least 10 hits~\cite{Alver:2008aq,STAR:2012och}. 
 In this analysis, tracks with at least $15$ TPC space points and Distance of Closest Approach (DCA) to the primary vertex of less than 3~cm were used.  We accept tracks with transverse momentum $0.2<p_{\rm T}<4$~\GeVc. 
Events are chosen with vertex positions within $\pm 40$~cm from the TPC center (along the beam direction), and within $\pm 2$~cm in the radial direction relative to the center of the TPC.

The two- and multi-particle cumulant methods are employed for our correlation analysis. The framework for the cumulant method is described in Refs.~\cite{Bilandzic:2010jr,Bilandzic:2013kga}; its extension to the case of subevents is also described in Refs.~\cite{Gajdosova:2017fsc,Jia:2017hbm}. 
Here, the two- and multi-particle correlations were formed using the two-subevents cumulant technique~\cite{Jia:2017hbm},  with  $\Delta\eta~=~\eta_{1}-\eta_{2}~ > 0.7$  between the subevents $\textit{A}$ and $\textit{B}$ (\textit{i.e.}, $\eta_{A}~ > 0.35$ and $\eta_{B}~ < -0.35$).  The use of the two-subevents technique serves to reduce the nonflow correlations~\cite{Magdy:2020bhd}. The two-  and multi-particle correlations are given as:
\begin{eqnarray}\label{eq:2-1}
v^{\rm Inclusive}_{k}                        &=&  \langle  \langle \cos (k (\varphi^{A}_{1} -  \varphi^{B}_{2} )) \rangle \rangle^{1/2},
\end{eqnarray}
\begin{eqnarray}\label{eq:2-2}
C_{k,nm}                                          &=&   \langle \langle \cos ( k \varphi_{1}^{A} - n \varphi_{2}^{B} -  m \varphi_{3}^{B}) \rangle \rangle ,
\end{eqnarray}
 {\small
\begin{eqnarray}\label{eq:2-3}
\langle v_{n}^{2} v_{m}^{2}  \rangle &=& \langle \langle \cos ( n \varphi^{A}_{1} + m \varphi^{A}_{2} -  n \varphi^{B}_{3} -  m \varphi^{B}_{4}) \rangle \rangle, \nonumber \\ 
\end{eqnarray}
}
where $\langle \langle \, \rangle \rangle$ denotes the average over all particles in a single event and a subsequent  average over all events, $\textit{k} =\textit{n}+\textit{m}$, $\textit{n} = 2$, $\textit{m} = 2$ or $3$, and $\varphi_{i}$ is the azimuthal angle of the momentum of the $\textit{i}^{th}$ particle. 

Using  Eqs.~(\ref{eq:2-1})-(\ref{eq:2-3}), the mode-coupled contributions to $v_{k}$, assuming factorization, can be expressed as~\cite{Yan:2015jma,Bhalerao:2013ina}:
\begin{eqnarray}\label{eq:2-4}
\textit{v}_{k}^{\rm MC} &=&  \frac{C_{k,nm}} {\sqrt{\langle \textit{v}_n^2 \textit{v}_m^2 \rangle}},  \nonumber \\  
                                     &\sim & \langle \textit{v}_{k} \, \cos (k \Psi_{k} - n\Psi_{n} - m\Psi_{m}) \rangle,
\end{eqnarray}
and the linear contribution to $v_{k}$ is given by:
\begin{eqnarray}\label{eq:2-5}
v_{k}^{\rm Linear}  &=& \sqrt{ (v^{\rm Inclusive}_{k})^{\,2} - (v^{\rm MC}_{k})^{\,2}  }.
\end{eqnarray}
Equation (\ref{eq:2-5}) assumes that the linear and mode-coupled contributions to $v_{k}$ are independent~\cite{Yan:2015jma,Magdy:2020bhd}.
The ratio of the mode-coupled contribution to the inclusive $v_{k}$ also gives an estimate of the correlation $\rho_{k,nm}$ between flow symmetry planes of order $n$ and $m$~\cite{Acharya:2017zfg};
\begin{eqnarray}\label{eq:2-6}
\rho_{k,nm} &=& \frac{v^{\rm MC}_{k}}{v^{\rm Inclusive}_{k}} ,  \nonumber \\ 
                   &\approx & \langle  \cos (k \Psi_{k} - n \Psi_{n} - m \Psi_{m}) \rangle.
\end{eqnarray}
The mode-coupled response coefficients, $\chi_{k,nm}$, which quantify the contributions of the coupling to the the higher-order anisotropic flow harmonics, are defined as:
\begin{eqnarray}\label{eq:2-7}
\chi_{k,nm} &=& \frac{v^{\rm MC}_{k}} {\sqrt{\langle  v_{n}^{2} \, v_{m}^{2} \rangle}}.
\end{eqnarray}
The normalized symmetric cumulants, $\mathrm{NSC (n,m)}$, from the standard cumulants method 
~\cite{Bilandzic:2010jr,Bilandzic:2013kga}  are given as:
\begin{eqnarray}
\mathrm{SC (n,m)}   &=& \langle \langle \cos ( n \varphi_{1} + m \varphi_{2} -  n \varphi_{3} -  m \varphi_{4}) \rangle \rangle    \nonumber \\
               &-& \langle  \langle \cos(n (\varphi_{1} -  \varphi_{2})) \rangle \rangle   \nonumber \\
               & & \langle  \langle \cos(m (\varphi_{1} -  \varphi_{2})) \rangle \rangle   \label{eq:2-8} \\
               \mathrm{NSC (n,m)} &=&  \dfrac{\mathrm{SC (n,m)}}{ \left( v^{\rm Inclusive}_{n}\right)^{2} \left( v^{\rm Inclusive}_{m}\right)^{2}}, \label{eq:2-9}
\end{eqnarray}
with the condition that $\textit{m} \neq \textit{n}$ and $\textit{n}$ and $\textit{m}$ are positive integers.
The $p_T$-integrated measurements for $\textit{k} =\textit{n}+\textit{m}$, $\textit{n} = 2$, $\textit{m} = 2$ and $3$ were performed as a function of centrality for each beam energy.

The systematic uncertainties of the presented measurements are obtained from variations in the analysis cuts for event selection, track selection and non-flow suppression;
(I) event selection was varied via cuts on the vertex positions determined in the TPC along the beam 
direction, $-40$ to $0$~cm or $0$ to $40$~cm instead of the nominal value of $\pm 40$~cm. 
(II) Track selection was varied by (a) reducing the DCA from its nominal value of 3~cm to 2~cm, and (b) increasing the number of TPC space points from greater than $15$ points to more than $20$ points. (III) The pseudorapidity gap, $\Delta\eta$ for the track pairs, used to mitigate the non-flow effects due to resonance decays, Bose-Einstein correlations, and the fragments of individual jets, was varied from $\Delta\eta = 0.6$ to $\Delta\eta = 0.8$. 
 \begin{figure}[H] 
  \vskip -0.2cm
 \centering{
 \includegraphics[width=0.99\linewidth, angle=0]{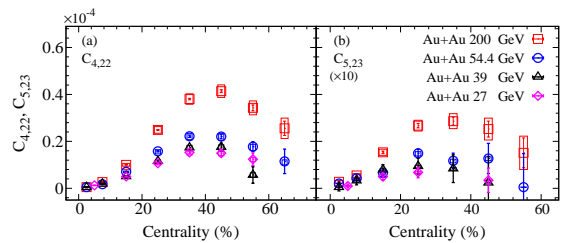}
  \vskip -0.3cm
\caption{Comparison of the \pt integrated three-particle correlators, $C_{4,22}$ (a) and $C_{5,23}$ (b), for Au+Au collisions at \roots = 54.4, 39 and 27~GeV, obtained with the  two-subevents cumulant method. 
The $C_{4,22}$ and $C_{5,23}$ measurements for Au+Au at \roots = 200~GeV are taken from Ref.~\citep{Adam:2020ymj}.
\label{Fig:1} }}
 \vskip -0.0cm
\end{figure}
Table~\ref{tab:1} gives a summary of these systematic uncertainty estimates. The overall systematic uncertainty, assuming independent sources, was evaluated via a quadrature sum of the uncertainties resulting from the respective cut variations. They range from 4\% to 10\% from central to peripheral collisions. The overall systematic uncertainties are shown as open boxes in the figures. Statistical uncertainties are shown as vertical lines.
\begin{table}[H]
\begin{center}
 \begin{tabular}{|c|c|c|}
 \hline
  Quantities        &          Minimum value        &                   Maximum value                      \\
 \hline
  Event                          &              2\%                   &                      5\%                                   \\
 \hline 
 Track                           &              3\%                   &                      7\%                                   \\
 \hline
 $\Delta\eta$                    &              2\%                   &                      7\%                                    \\
 \hline
\end{tabular} 
 \vskip -0.2cm
\caption{Summary of the estimated systematic uncertainty contributions (see text).}
\label{tab:1}
\end{center}
 \vskip -0.8cm
\end{table}

%
\end{multicols}
 \begin{figure*}[t]
  \vskip -0.4cm
 \centering{
\includegraphics[width=0.75\linewidth,angle=0]{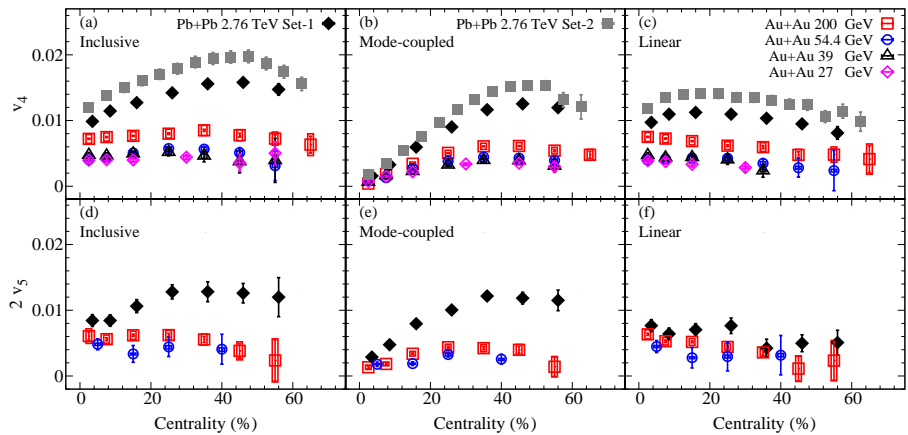}
\vskip -0.5cm
\caption{Comparison of the inclusive ((a) and (d)), mode-coupled ((b) and (e)) and linear ((c) and (f)) higher-order flow harmonics \fourth and \fifth obtained with the two-subevents cumulant method, as a function of centrality in the $p_{\rm T}$ range $0.2 - 4.0$~\GeVc~ for Au+Au collisions at \roots = 54.4, 39 and 27~GeV. The \fourth and \fifth measurements of  \roots = 200~GeV are taken from Ref.~\citep{Adam:2020ymj}.
{\color{black} The solid points indicate LHC measurements for $0.2$ $<$ $p_{\rm T}$ $<$ $5.0$~\GeVc~ from the ALICE experiment (set-1)~\cite{Acharya:2017zfg} and for $0.5$ $<$ $p_{\rm T}$ $<$ $2.0$~\GeVc~ from the ATLAS experiment (set-2)~\citep{Aad:2015lwa} for Pb+Pb collisions at \roots = 2.76~TeV.}
 \label{Fig:2} }}
 \vskip -0.2cm
\end{figure*}

\begin{multicols}{2}

Figure~\ref{Fig:1} compares the centrality dependence of the $C_{4,22}$ and $C_{5,23}$ coefficients for $0.2 < p_{\rm T} <4.0$~\GeVc in Au+Au collisions at \roots = 200, 54.4, 39 and 27~GeV. The coefficients show similar centrality dependent patterns and magnitudes that decrease with beam energy. These dependencies suggests that $C_{4,22}$ and $C_{5,23}$ are sensitive to the initial-state eccentricity and the change in viscous attenuation with beam energy. The latter could result from both a change in the charge particle multiplicity and  $\eta/s(\mu_B, T)$~\cite{Lacey:2013qua,Karpenko:2015xea} with beam energy. Thus, detailed model comparisons to the centrality and beam energy dependence of $C_{4,22}$ and $C_{5,23}$ could serve as an additional constraint for precision extraction of \etas\cite{Magdy:2021sba}.


The $v^{\rm Inclusive}_{k}$, $C_{4,22}$ and $C_{5,23}$ coefficients were used to extract $\textit{v}_{k}^{\rm MC}$, $v_{k}^{\rm Linear}$, $\rho_{k,nm}$, $\chi_{k,nm}$, and $\mathrm{NSC (n,m)}$ (cf. Eqs.~\ref{eq:2-4} -- \ref{eq:2-9}) to home in on further constraints for the initial- and final-states respectively. The centrality dependence of $v^{\rm Inclusive}_{k}$ ((a) and (d)), $v_{k}^{\rm Linear}$ ((b) and (e)), and $\textit{v}_{k}^{\rm MC}$ ((c) and (f)) $\textit{v}_{4,5}$ coefficients are shown for several beam energies in Fig.~\ref{Fig:2}. The mode-coupled coefficients ((b) and (e)) indicate a much stronger increase with centrality than that for the linear coefficients ((c) and (f)), suggesting that the $v_{k}^{\rm Linear}$  coefficients are subject to much larger viscous attenuation than the $\textit{v}_{k}^{\rm MC}$ coefficients; note that ${\varepsilon_{k}^{\rm MC}}$ and $\varepsilon_{k}^{\rm Linear}$ increase with centrality. The $\textit{v}_{k}^{\rm MC}$ and $v_{k}^{\rm Linear}$ coefficients for Au+Au collisions also indicate a relatively weak dependence on beam energy, suggesting that the viscous attenuation and the eccentricity are weak functions of the beam energy (cf. Eq.~\ref{eq:1-2}) especially for the energy span \roots = 27 - 54.4~GeV. 
{\color{black} The LHC measurements (set-1~\cite{Acharya:2017zfg}, or ALICE measurements for $0.2 < p_{\rm T} <5.0$ ~\GeVc~ and  $|\eta|<0.8$, and set-2~\cite{Aad:2015lwa}, or ATLAS measurements for $p_{\rm T} > 0.5$ ~\GeVc~ and  $|\eta|<\,2.5$) (panels (a)--(f))}
 show patterns that are similar to those for Au+Au collisions, albeit with magnitudes that are much larger, implying a more  sizable dependence on beam energy from RHIC to LHC energies \cite{Lacey:2013qua,Magdy:2021sba}. The difference between the magnitudes for the set-1 and set-2 LHC measurements reflects the dependence of these coefficients on $\langle p_{T} \rangle$. 
Note however, that the $\left< p_T \right>$ is a weak function of the RHIC beam energy range of interest in this work~\cite{STAR:2017sal}. 
These beam energy and centrality dependencies can be used to further constrain theoretical models.

 \begin{figure}[H]
 \centering{
  \vskip -0.2cm
 \includegraphics[width=0.99\linewidth, angle=0]{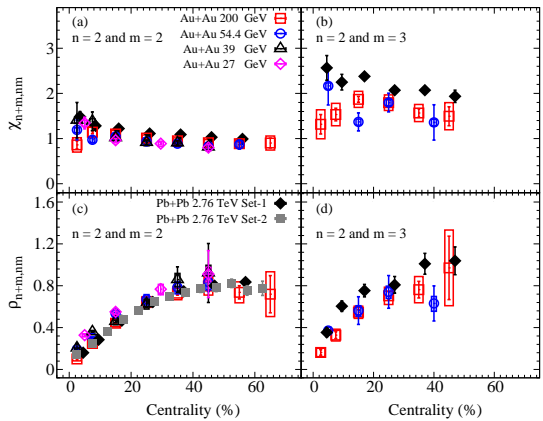}
 \vskip -0.5cm
 \caption{Comparison of the $\chi_{n+m,nm}$ ((a) and (c)) and $\rho_{n+m,nm}$ ((b) and (d)) obtained with the two-subevents cumulant method, as a function of centrality in the $p_{\rm T}$ range $0.2 - 4.0$~\GeVc~ for Au+Au collisions at \roots = 54.4, 39 and 27~GeV. The $\chi_{n+m,nm}$ and $\rho_{n+m,nm}$ at \roots = 200~GeV are taken from Ref.~\citep{Adam:2020ymj}.
 The solid points are the LHC  measurements for Pb+Pb collisions at \roots = 2.76~TeV~set-1~\cite{Acharya:2017zfg} and set-2~\citep{Aad:2015lwa}.
 \label{Fig:3} }}
 \vskip -0.2cm
 \end{figure}
 \begin{figure}[H]
  \vskip -0.2cm
 \centering{
 \includegraphics[width=0.7\linewidth, angle=0]{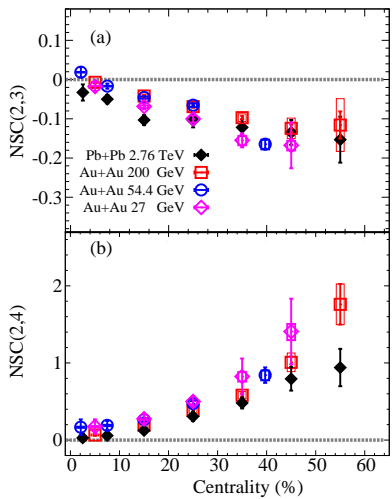}
 \vskip -0.7cm
 \caption{Comparison of $\mathrm{NSC (2,3)}$ (a) and $\mathrm{NSC (2,4)}$ (b) using the standard cumulant method  as a function of centrality in the $p_{\rm T}$ range 0.2--4.0~\GeVc~ for Au+Au collisions at \roots = 54.4, 39 and 27~GeV. 
 The $\mathrm{NSC (2,3)}$ and $\mathrm{NSC (2,4)}$ at \roots = 200~GeV are taken from Ref.~\citep{STAR:2018fpo}.
 The solid diamonds indicate LHC  measurements for the $p_{\rm T}$ range from 0.2--5.0~\GeVc~ for Pb+Pb collisions at \roots = 2.76~TeV~\cite{ALICE:2016kpq}.
 \label{Fig:4} }}
 \vskip -0.2cm
 \end{figure}

The centrality dependence of the mode-coupled response coefficients $\chi_{k,nm}$ ($n=2$ and $m=2$ and $3$) for Au+Au (\roots = 200, 54.4, 39 and 27~GeV) and Pb+Pb collisions (\roots = 2.76~TeV)~\cite{Acharya:2017zfg}  are compared in Figs. \ref{Fig:3} (a) and (b). Results demonstrate a weak dependence on centrality and beam energy,  confirming that (I) the mode-coupled $\textit{v}_{4,5}$ coefficients are dominated by the correlations from the lower-order flow harmonics and (II)  $\chi_{k,nm}$ is weakly sensitive to the viscous effects ($\etas/s$)~\cite{Lacey:2013qua,Magdy:2021sba} and hence, more sensitive to the initial-state effects.

Figure~\ref{Fig:3} (c) and (d) compares the centrality dependence of the $\rho_{k,nm}$ coefficients  for Au+Au collisions  (\roots = 200, 54.4, 39 and 27~GeV) and Pb+Pb collisions (\roots = 2.76 TeV)~\cite{Acharya:2017zfg}. Within the indicated uncertainties, they indicate a strong centrality dependence and a relatively weak dependence on beam energy. These characteristic dependencies suggest that $\rho_{k,nm}$ can provide an additional constraint for the beam energy dependence of the viscous effects ($\eta/s$)~\cite{Lacey:2013qua,Magdy:2021sba} and could be used to discern different initial-state models~\cite{Magdy:2021sba}.

Figure~\ref{Fig:4} summarizes the results for the $\mathrm{NSC (n,m)}$ that reflect the strength of the correlation/anti-correlation between the $\textit{v}_{n}$ and $\textit{v}_{m}$ flow harmonics. 
Figs.~\ref{Fig:4}(a) and (b) show the $\mathrm{NSC (2,3)}$ and $\mathrm{NSC (2,4)}$ respectively, for $0.2 < p_{\rm T} < 4.0$~\GeVc~ in Au+Au collisions at \roots = 200, 54.4 and 27~GeV and the corresponding LHC measurements~\cite{ALICE:2016kpq}.
The $\mathrm{NSC (2,3)}$ coefficients indicate an anti-correlation (negative values)~\cite{Giacalone:2016afq,Zhou:2016eiz} between \second and \third, as expected from the known anti-correlation between $\mathrm{\varepsilon_{2}}$ and $\mathrm{\varepsilon_{3}}$.  In contrast, the $\mathrm{NSC (2,4)}$ coefficients indicate a correlation between \second and \fourth consistent with the mode-coupled correlations between $\mathrm{\varepsilon_{2}}$ and $\mathrm{\varepsilon_{4}}$. Within the uncertainties, the weak beam energy dependence further indicates that $\mathrm{NSC (2,3)}$ and $\mathrm{NSC (2,4)}$ are less sensitive to the effects of viscous attenuation \cite{Lacey:2013qua} and could set a constraint on the initial-state eccentricity correlations.  

In summary, we have presented new \pt-integrated measurements of the charge-inclusive, linear and mode-coupled 
contributions to the higher-order anisotropic flow coefficients $\textit{v}_{4,5}$, mode-coupled response 
coefficients $\chi_{k,nm}$, correlations of the event plane angles $\rho_{k,nm}$ and  normalized symmetric cumulant $\mathrm{NSC (2,3)}$ and $\mathrm{NSC (2,4)}$, for  Au+Au collisions at \roots = 200, 54.4, 39 and 27~GeV. Our measurements are compared with similar LHC measurements  for Pb+Pb collisions at \roots = 2.76~TeV.
For all presented energies, the mode-coupled $\textit{v}_{4,5}$  measurements indicate a large centrality dependence. In contrast, the linear $\textit{v}_{4,5}$, which dominates the central collisions, displays a weak centrality dependence. The $\textit{v}_{4,5}$  measurements show a  beam energy dependence which reflects the sensitivity to  $\eta / \textit{s}$.
The dimensionless coefficients $\chi_{k,nm}$, $\rho_{k,nm}$, $\mathrm{NSC (2,3)}$ and $\mathrm{NSC (2,4)}$ show magnitudes and trends which are approximately beam energy independent, suggesting that the measured dimensionless quantities are dominated by initial-state effects.
These results should prove invaluable to theoretical efforts which seek simultaneous constraints for $\frac{\eta}{s}(T, \mu_B)$ and the initial-state.

\section*{Acknowledgments}
We thank the RHIC Operations Group and RCF at BNL, the NERSC Center at LBNL, and the Open Science Grid consortium for providing resources and support.  This work was supported in part by the Office of Nuclear Physics within the U.S. DOE Office of Science, the U.S. National Science Foundation, National Natural Science Foundation of China, Chinese Academy of Science, the Ministry of Science and Technology of China and the Chinese Ministry of Education, the Higher Education Sprout Project by Ministry of Education at NCKU, the National Research Foundation of Korea, Czech Science Foundation and Ministry of Education, Youth and Sports of the Czech Republic, Hungarian National Research, Development and Innovation Office, New National Excellency Programme of the Hungarian Ministry of Human Capacities, Department of Atomic Energy and Department of Science and Technology of the Government of India, the National Science Centre and WUT ID-UB of Poland, the Ministry of Science, Education and Sports of the Republic of Croatia, German Bundesministerium f\"ur Bildung, Wissenschaft, Forschung and Technologie (BMBF), Helmholtz Association, Ministry of Education, Culture, Sports, Science, and Technology (MEXT) and Japan Society for the Promotion of Science (JSPS).

%

\bibliographystyle{elsarticle-num}
\bibliography{ref}

\end{multicols}

\end{document}